\input feynman

\documentstyle[12pt]{article}

\advance\textheight by \topskip
\begin{document}
\tolerance=100000
%
%minore/uguale
\def\Ord{\buildrel{\scriptscriptstyle <}\over{\scriptscriptstyle\sim}}
%maggiore/uguale
\def\OOrd{\buildrel{\scriptscriptstyle >}\over{\scriptscriptstyle\sim}}

\thispagestyle{empty}
\setcounter{page}{0}

\begin{flushright}
{\large DFTT 44/93}\\
{\rm September 1993\hspace*{.5 truecm}}\\
\end{flushright}

\vspace*{\fill}

\begin{center}
{\Large \bf Trying to catch the elusive $A^0$ and more.
Reactions initiated by $b$--quarks and the
Higgs sector of the ${\cal MSSM}$\footnote{ Work supported in part by Ministero
dell' Universit\`a e della Ricerca Scientifica.\\[4. mm]
Mail address: V. Giuria 1, 10125 Torino, Italy.\\
e-mail: ballestrero,maina,moretti@to.infn.it}}\\[2.cm]
{\large Alessandro Ballestrero, Ezio Maina, Stefano Moretti}\\[.3 cm]
{\it Dipartimento di Fisica Teorica, Universit\`a di Torino, Italy}\\
{\it and INFN, Sezione di Torino, Italy.}\\[1cm]
{\large Corrado Pistarino}\\[.3 cm]
{\it Dipartimento di Fisica Teorica, Universit\`a di Torino, Italy}\\[1cm]
\end{center}

\vspace*{\fill}

\begin{abstract}
{\normalsize
\noindent
We study the cross sections for
the production of a neutral, intermediate mass Higgs boson in the processes
$pp\rightarrow tq'\Phi$, $pp\rightarrow tW^-\Phi$
and $pp\rightarrow bZ^0\Phi$ in the Minimal Supersymmetric Standard Model
($\Phi=H^0,h^0$ and $A^0$) at Supercollider energies.
The additional heavy particles ($t$, $W$, $Z$) in the final state
can be used for tagging purposes, increasing the signal to background ratio.
These reactions are dominated by $bq$ and $bg$ fusion.
Their relevance for
Higgs particle searches is discussed taking into account the expected
efficiencies and purities for $b$--tagging.
We find that, for tan$\beta = 30$, the cross sections
for $pp\rightarrow bZ^0\Phi$ are larger than 14 $pb$,
over the whole intemediate range of $M_{A^0}$, for $A^0$ and at least one of
the other two Higgses. Therefore this reaction is an excellent candidate for
the discovery of one or more $\cal MSSM$ Higgs particles.}
\end{abstract}

\vspace*{\fill}

\newpage
\subsection*{Introduction}

Both in the Standard Model (${\cal SM}$) and in the Minimal
Supersymmetric Standard Model (${\cal MSSM}$) the Higgs
mechanism is assumed, after the spontaneous symmetry
breaking of the $SU(2)\times U(1)$ gauge group, to give
masses to gauge bosons, to fermions and, in the latter,
to their supersymmetric partners.\par
While in the ${\cal SM}$ a doublet of complex scalar
fields is sufficient to induce the symmetry breaking, in the ${\cal MSSM}$
this requires two doublets.\par
Of the initial degrees of freedom, three are employed
to give a longitudinal polarization to the weak gauge bosons
$Z^0$ and $W^\pm$; the remaining ones, one for the ${\cal SM}$ and
five for the ${\cal MSSM}$, appear in the theory as interacting
scalar particles: the Higgs bosons.\par
The ${\cal SM}$ Higgs $\phi$ is a $CP$--even neutral particle; among
the ${\cal MSSM}$ Higgses three are neutral, the $CP$--even ones $H^0$
and $h^0$ and the $CP$--odd one $A^0$, and two charged, the $H^\pm$'s.
The three neutral Higgs states of the ${\cal MSSM}$ will be collectively
indicated by $\Phi$.\par
Unitarity of the theory imposes a ${\cal SM}$ upper limit \cite{unitsm}
\begin{equation}
M_\phi\Ord \left({{8\sqrt 2\pi}\over{3G_F}}\right)^{1/2}\sim
1~{\rm TeV},
\end{equation}
where $G_F$ is the Fermi electroweak constant. The analysis
is more complicated in models with an extended Higgs sector \cite{guide},
such as the ${\cal MSSM}$, but similar arguments indicate that
at least one neutral scalar must have mass below $\sim 1$
TeV \cite{unitmssm1,unitmssm2}.\par
In the simplest version of the ${\cal MSSM}$ all Higgs masses are predicted
at tree level as a function of tan$\beta$, the ratio of the vacuum
expectation value of the two doublets, and $M_{A^0}$, the mass of the
$CP$--odd state.
At one--loop these predictions are substantially modified and an additional
dependence on the top mass $m_t$ and on the common squark mass
$m_{\tilde t}$ is introduced.
One has \cite{wisco1}:
\begin{eqnarray}\label{m1}
M^{2}_{h^0,H^0}& = & \frac{1}{2}[M^{2}_{A^0} + M_{Z^0}^{2} +
\epsilon/\sin^{2}\beta] \nonumber \\
           &   & \pm \left\{ [ (M^{2}_{A^0} - M^{2}_{Z^0})\cos2\beta +
\epsilon/\sin^{2}\beta]^{2}
                 +(M^{2}_{A^0} + M^{2}_{Z^0})^{2}{\rm sin}^{2}2\beta
\right\}^{1/2},
\end{eqnarray}
where
\begin{equation}\label{m2}
\epsilon = \frac{3e^{2}}{8\pi^{2} M^{2}_{W}{\rm sin}^2\theta_W}m_{t}^{4} {\rm
ln}\left( 1 +
\frac{{m}^{2}_{\tilde t}}{m_{t}^{2}} \right).
\end{equation}
The squark mass scale ${m}_{\tilde t}$ is expected to be of the order of
1 TeV.
The mixing angle $\alpha$ in the $CP$--even sector, which together with
$\beta$ determines all couplings of the ${\cal MSSM}$ Higgses (table I),
is defined by
\begin{equation}\label{m3}
\tan 2\alpha = \frac{(M_{A^0}^{2} + M_{Z^0}^{2}){\rm sin}2\beta}{(M_{A^0}^{2} -
M_{Z^0}^{2})
{\rm cos2}\beta + \epsilon/{\rm sin}^{2}\beta}.
\end{equation}
As it will be discussed in more detail later on, we are mainly interested in
the intermediate range mass for $A^0$,
 which is the region in parameter space which is the most difficult
to explore experimentally.
For relatively large values of
tan$\beta$, two different regimes can be distinguished depending on whether
$M_{A^0}$
is smaller or larger than a treshold value of 100--130 GeV.
For lower $M_{A^0}$, $M_{H^0}\approx 110$ GeV while $M_{h^0}\approx M_{A^0}$
and $\alpha \approx -90^\circ$,
while for larger $M_{A^0}$ the role of $h^0$ and $H^0$ are exchanged
and $\alpha \approx 0^\circ$.
The region in between these two regimes, where all the couplings of
$h^0$ and $H^0$ to quarks are simultaneously suppressed, corresponds to the
intermediate mass region for $A^0$.\par
Only lower limits on the Higgs masses have been extracted at present
colliders. LEP experiments,
from the results of searches for $Z^{0*}\phi$ events, derive a bound
\begin{equation}
M_\phi\OOrd 62.5~{\rm GeV},
\end{equation}
for the ${\cal SM}$ Higgs \cite{aleph}.
Using the reactions $e^+e^-\rightarrow Z^{0*}h^0$ and
$e^+e^-\rightarrow h^0A^0$, the lower limits on ${\cal MSSM}$
neutral Higgses are presently
\begin{equation}
M_{h^0}\OOrd 44.5~{\rm GeV}\quad\quad{\rm and}\quad\quad M_{A^0}\OOrd 45~{\rm
GeV},
\end{equation}
for the typical choice of parameters $m_t=140$ GeV and $m_{\tilde t}
=1$ TeV \cite{aleph}.\par
Extensive studies have been carried out on the detectability of
a Higgs particle by the next generation of high energy
colliders \cite{guide,hreview1,hreview2,hreview3,kz,wisco2}.
The region $M_\Phi < 80$ GeV will be studied at LEP II.
For $M_{A^0} < $80--90 GeV one or both of the two processes
$e^+e^-\rightarrow Z^{0*}h^0$ and $e^+e^-\rightarrow h^0A^0$ will be
discovered.
Higgses with larger masses will be searched for
at $pp$ colliders like LHC ($\sqrt s=16$ TeV) and SSC ($\sqrt s=40$ TeV).
The intermediate--mass range 80 GeV $\Ord M_\Phi\Ord 130$ GeV is the
most difficult one.
In this range a Higgs boson $\Phi$ mainly
decays to $b\bar b$ pairs, both in the ${\cal SM}$ and,
for a large choice of parameters, in the
${\cal MSSM}$. Because of the huge QCD background, its
detection in this channel results very difficult.
The discovery of a Higgs in the two--photon inclusive mode requires
an extremely good photon--pair mass resolution.
However, it has been established that
the associated production of a ${\cal SM}$ Higgs
$\phi$ with a $W^\pm$ boson \cite{gny,wh}
or a $t\bar t$ \cite{rwnz,tth} pair,
followed by the decays $\phi\rightarrow\gamma\gamma$
and $W\rightarrow\ell\nu$, can be revealed with the diphoton
mass resolution expected from SSC/LHC detectors \cite{gamgam}.
Requiring the presence of a highly energetic and isolated lepton in
the final state is a very effective method to enhance the
signal to background ratio.
The branching ratio to $\gamma\gamma$ depends crucially on the
Higgs coupling to the top quark. In the $\cal MSSM$ this mode
can be exploited for the discovery of $H^0$
for 80 GeV $< M_{A^0} <$ 100 GeV
and for the discovery of $h^0$ for $M_{A^0}>$ 170 GeV.
For $M_\phi > 130$ GeV the four--lepton mode guarantees, in general,
the detection of the $\cal SM$ Higgs.
In the $\cal MSSM$ this decay channel is only useful for the $H^0$ for
moderate tan$\beta$ and 100 GeV $< M_{A^0} <$ 300 GeV.\par
Recently, it has been suggested \cite{ucd} that with the $b$--tagging
capabilities of SSC experiments, and possibly LHC ones if the higher
luminosity and larger number of tracks per event can successfully
be dealt with, it may be possible to detect a
Higgs boson in the $t\bar t\Phi$ production channel, requiring one $t$
to decay semileptonically and using the hadronic decay
$\Phi\rightarrow b\bar b$.
This idea could be used both for the ${\cal SM}$
Higgs $\phi$ and, over a wide range of the parameter space,
for at least one of the ${\cal MSSM}$ Higgses $h^0$ or $H^0$.
In particular this would close the ``window of inobservability''
which remained in previous analysis for
100 GeV $< M_{A^0} <$ 170 GeV and tan$\beta$
larger than about 2.
This channel is useless for $A^0$ because
its coupling to $t$--quarks is suppressed by 1/tan$\beta$.
It would be interesting to know whether vertex--tagging of the
$b$'s can disentangle the rather large signal from  $gg \rightarrow b\bar b
A^0$
\cite{kz} from the enormous $b\bar b b\bar b$ QCD background.\par
In this paper we study the following reactions:
\begin{equation}\label{e1}
b+q(\bar q)\longrightarrow t+q'(\bar q')+\Phi,
\end{equation}
\begin{equation}\label{e2}
b+g\longrightarrow t+W^-+\Phi,
\end{equation}
\begin{equation}\label{e3}
b+g\longrightarrow b+Z^0+\Phi.
\end{equation}
with the purpose to extend the range in parameter space for
which more than one $\cal MSSM$ Higgs can be detected, and in particular to
enlarge the regions of observability of $H^0$ and $A^0$.
All these reactions
\begin{enumerate}
\item contain $b$--quarks and therefore may be enhanced for large
     tan$\beta$;
\item have additional heavy particles $t, W, Z$ in the final state
      which can produce highly energetic and isolated leptons.
      This could be particularly useful at LHC, somewhat relaxing
      the requirements on $b$--tagging devices by decreasing the trigger
      rate and drastically reducing the background.
\end{enumerate}
The first process (\ref{e1})
is the supersymmetric version of the reaction studied in
\cite{tbh} in the $\cal SM$. It can be interpreted as the dominant
contribution to $gq\rightarrow t \bar b q'\Phi$ corresponding to
the gluon which splits into a collinear
$b\bar b$ pair.
The cross section depends on the Higgs coupling to $b$ and $t$--quarks
and to vector bosons; moreover it is known that there are large
cancellations among different diagrams, therefore it is impossible
to obtain the $\cal MSSM$ cross section simply multiplying the $\cal SM$
result by an overall factor.\par
The second one (\ref{e2}), again if read as a shorthand for
$gg\rightarrow  \bar b W^- t\Phi$, represents the contribution to
$gg \rightarrow b \bar b W^- W^+\Phi$ which does not proceed through
$gg\rightarrow  t \bar t \Phi$ and consequent
$t$ decay. Therefore it is expected to be a small
correction to $gg\rightarrow  t \bar t \Phi$ whenever this reaction
is relatively large. It could however be important in regions in which
$t \bar t \Phi$ production is suppressed, in particular
for $A^0$.\par
The last reaction (\ref{e3}) is described at tree level by the same set
of diagrams that describes (\ref{e2}), with a $Z^0$ replacing the $W^-$
boson and consequently with a $b$--quark in place of a $t$--quark in
the final state. The lighter mass of the $b$ makes this reaction
kinematically favored in comparison to (\ref{e2}). The branching ratio of the
$Z^0$ to light leptons is smaller than the corresponding one for $W$'s,
but the signature is much cleaner and does not suffer from the huge
background from $top$ production.
The presence of a $Z^0$ has the additional advantage of allowing a very good
measurement of the position of the primary vertex and therefore
facilitates the search for secondary vertices.\par
In the following section, we
give some details of the calculation
and the values adopted for the various parameters.
Section III is devoted to a discussion of the obtained results.
Conclusions are in section IV.

\subsection*{Calculation}

The Feynman diagrams describing at tree--level and in the unitary gauge
the reaction (\ref{e1})
are shown in fig.1. Some of the ones corresponding to  (\ref{e2})--(\ref{e3})
are
depicted in fig.2.\par
The diagrams with a direct coupling of $\Phi$
to the light quark line have been neglected, since they
vanish in the massless limit.
In fig.2  five diagrams are not shown. Four can be obtained
obtained exchanging the attachement of the vector bosons in (1) to (4),
the fifth exchanging the gluon and the higgs attachements in (5).\par
For a ${\cal SM}$ Higgs only the first three diagrams of fig.1 and
the first four of fig.2
contribute; in the ${\cal MSSM}$ case,
for a $CP$--odd Higgs boson ($\Phi=A^0$), the diagrams with the
coupling $VVA^0$, where $V=W^\pm,Z^0$, vanish.\par
Moreover, for the process (\ref{e1}) there are
additional contributions
from diagrams containing a massless antiquark line which are
included in our results.\par
We have evaluated the matrix elements in different ways.
All amplitudes have been calculated,
using spinor techniques \cite{ks,mana,noi,hz}, in two different gauges,
the unitary and the Feynman one, and checked for gauge
and BRST invariance \cite{brs1,brs2,brs3}. For process (\ref{e1}) we
have also computed the cross section using the time--honored
trace method.\par
Then, the matrix elements have been numerically integrated over
phase space using VEGAS \cite{vegas}.\par
For the electroweak parameters we have chosen sin$^2\theta_W=0.2325$
and $\alpha_{em}=1/128$, with the masses $M_{Z^0} = 91.173$ GeV and
$M_W=M_{Z^0}$cos$\theta_W$.
For the quark masses  the values
$m_t=150$ GeV and  $m_b=5$ GeV have been used throughout, while
$u,d,s$ and $c$ quarks have been considered massless.\par
The strong coupling constant $\alpha_s$ and the parton distribution
functions have been consistently evaluated at a scale
equal to the subprocess invariant mass.
We have used the one loop expression for $\alpha_s$,
with $\Lambda_{QCD}=150$ MeV and five active flavors.\par
In all of the calculations
the structure function set HMRSB has been used \cite{hmrsb}.
Changing the scale and/or distribution function choice
should not affect our predictions by more than a factor of two.\par
We have analyzed  the mass
range
50 GeV $<$ $M_{A^0}$ $<$ 180 GeV for tan$\beta=2$, 15 and
30, adopting the one--loop expression (\ref{m1}) for the ${\cal MSSM}$
neutral Higgs masses.
For the ${\cal MSSM}$ charged Higgs masses we have maintained the
tree--level expression
\begin{equation}
M_{H^\pm}^2=M_{A^0}^2+M_W^2,
\end{equation}
since one--loop corrections are quite small if compared
with the corresponding ones for neutral Higgses \cite{bri1}.\par

\subsection*{Results}

Our results are presented in fig.3 through 6.
In order to assess their significance it is useful to
recall, as a reference point, that, in the regions in which detection
of $h^0$ or $H^0$ through
their tagged hadronic decay is possible \cite{ucd},
the cross section for $t\bar t h^0, H^0$ is about 5 $pb$.\par
In fig.3 we show
a number of cross sections for tan$\beta$ = 2. As expected they
are generally small.
The cross sections for $b+g\rightarrow t+W +h^0$ and
$b+g\rightarrow b+Z^0+h^0$ increase sharply at $M_{A^0} \approx 130$ GeV
and $M_{A^0} \approx 170$ GeV respectively. This is due to the onset
of the decay channel $H^\pm\rightarrow W^\pm h^0$ in the first case and
$A^0\rightarrow Z^0 h^0$ in the second. In order to check the
consistency of our results in the threshold regions we have also
estimated the
two cross sections in the narrow width approximation, assuming
diagram 5 in fig.2 to give the dominant contribution.
We have computed
the cross section for $g(p_1)+b(p_2)\rightarrow b(p_3)+A^0(p_4)$
with the following result:
\begin{eqnarray}\label{bgtoba}
\mid {\cal M}(bg\rightarrow bA)\mid^{2}_{\rm ave}& = &
                \frac{2\pi^2\alpha\alpha_s m^2_b \tan^2\beta }{3 M^2_W
\sin^2\theta_W}
                     \left[ \frac{2 m^2_b M^2_A}{(s-m^2_b)^2}
                              +\frac{2 m^2_b M^2_A}{(t-m^2_b)^2} \right. \\
           &   & \left. +\frac{2u (2 m^2_b - u)}{(s-m^2_b)(t-m^2_b)} +\frac{4
m^2_b - M^2_A - u}{(s-m^2_b)}
                 +\frac{4 m^2_b - M^2_A - u}{(t-m^2_b)}
                  \right] \nonumber
\end{eqnarray}
where $s =(p_1+p_2)^2$, $t =(p_1-p_3)^2$ and $u =(p_1-p_4)^2$.
The cross section for $b+g\rightarrow t+H^-$ has been taken
from ref. \cite{barnett}
while the branching ratios and the corresponding widths
have been derived from the
formulae given in \cite{guide}
using the one--loop--corrected masses. We obtain
$\sigma(b+g\rightarrow b+A^0\ +\ {\rm c.c.})\times
{\rm BR}(A^0\rightarrow Z^0 h^0) = .5 \ pb$
at $M_{A^0}= 180$ GeV and
$\sigma(b+g\rightarrow t+H^-\ +\ {\rm c.c.})\times
{\rm BR}(H^-\rightarrow W^- h^0) = 1.3 (.7) \ pb$
at $M_{A^0}= 140 (180)$ GeV in reasonable agreement with the full result.
Therefore we conclude that, for
tan$\beta$ =2, the processes we have examined
have cross sections at most of the order of 1 $pb$, even when new
decay channels open for intermediate--state particles.\par
In fig.4 we present the cross section for
$b+q(\bar q)\rightarrow t+q'(\bar q')+\Phi$ for large values of tan$\beta$.
The cross sections for $h^0$ and $H^0$
follow the trend of the respective coupling to the $b$, particularly
for tan$\beta$ = 30. In the $\cal SM$
there are large cancellations between diagrams (2) and (3) in fig.1.
In the supersymmetric case this cancellation is not substantially upset
because the factors which suppress the couplings of the two Higgses to the
$top$ and to the vector bosons are of similar magnitude throughout the
whole range in $M_{A^0}$ under consideration. Therefore the change of the
cross section reflects the change in the contribution of diagram (1).
In order to evaluate the number of events produced by the various
reactions one can adopt the SDC estimates of a 30\% efficiency
for single $b$--tagging. Then, assuming a factor of two reduction of the signal
for acceptance and kinematical cuts,
the probability of detecting three $b$'s together with
one high--$p_T$ lepton from $top$ or $W$ decay is about $2.7\times 10^{-3}$.
This gives 27 events per
standard SSC year ($L = 10^4 \quad pb^{-1}$) in the $t\Phi$ final state
for a cross section of 1 $pb$. The corresponding figure for the
$tW\Phi$ final state would be twice as large.
\par
The cross section for $bg\rightarrow tW\Phi$ is presented in fig.5.
This reaction results in a final state which is very similar
to $t\bar t \Phi$.
The higher luminosity of the gluon and the presence of a strong vertex
in place of an electroweak one do not compensate the effect of producing
an additional heavy particle and the rate is generally smaller, for a given
Higgs mass, than the corresponding rate for reaction (\ref{e1}) in
fig.4.
Summing the results in fig.4 and 5, at tan$\beta = 30$, one obtains for
$A^0$ a cross section which varies exponentially
from about 3 $pb$ at $M_{A^0} = 50$ GeV, to 2 $pb$ at $M_{A^0} = 100$ GeV,
to approximately .8 $pb$ at $M_{A^0} = 180$ GeV.
\par
We remark that for tan$\beta$ =15 and 30
the decay channel $H^\pm\rightarrow W^\pm h^0$ opens up at
$M_{A^0} \approx 180$ GeV. However the coupling of the $H^\pm$
to $h^0W^\pm$ is strongly suppressed at large tan$\beta$ and the
contribution from diagram 5 in fig.2 is negligible and the crossing
of the threshold gives no visible effect in the
cross sections for $b+g\rightarrow t+W +h^0$.\par
In fig. 6 we give the cross sections for $b+g\rightarrow b+Z^0+\Phi^0$,
$\Phi = h, H, A$ which are  by far the most relevant results we have obtained.
At least two of the three Higgses have cross sections larger than 14 $pb$
over the whole intemediate range of $M_{A^0}$ for tan$\beta = 30$.
As obvious the $A^0$ cross section for tan$\beta = 15$ is simply one fourth of
the
cross sections for tan$\beta = 30$. The same relationship between
the cross sections at the two values of tan$\beta$ also holds for
$h^0$ in the region $M_{A^0} < 100$ GeV and for $H^0$ in the
region $M_{A^0} > 120$ GeV, where the masses of the two
$CP$--even Higgs bosons are approximately independent of
$M_{A^0}$. Outside these two regions the cross sections for
$h^0$ and $H^0$ decrease rapidly. To get a feeling for the expected rates
we notice that taking into account the 6\% branching ratio
of $Z^0$ to light leptons, again assuming the SDC estimates for the
single $b$--tagging efficiency and the usual factor of two reduction
for acceptance, a cross section of 1 $pb$ corresponds to
27 events per SSC year, in which one $Z^0$ decays to
$\ell^+\ell^-$ and both $b$'s from
the Higgs decay are detected.
In this case the dominant background comes from $b\bar b Z^0$
production. In \cite{thesis} the total cross section for
$p\bar p \rightarrow b\bar b Z^0$ has been found to be
about 3.6 $nb$. Whether or not the $A^0$ can be detected in the
$b\bar b \ell^+\ell^-$ mode depends therefore on the $b\bar b$ mass--spectrum
of the background and on the detector mass resolution.
We have left this subject for future studies.
If all three $b$'s are required to be tagged, then one expects
about 9 events per SSC year for each $pb$ of cross section.
Possible backgrounds to this channel are
the irriducible one from $b\bar b b\bar b Z^0$ production, possibly with
a small contribution from $t\bar t b\bar b Z^0$, and $b\bar b Z^0+jets$
in which one $jet$ is misidentified as a $b$. Unfortunately,
to our knowledge, no estimate
for these processes is available  but it is very difficult
to imagine that they could be larger than the background to three $b$'s and
one high--$p_T$ lepton which has been studied in \cite{ucd}.
Therefore we believe that reaction (\ref{e3}) is a good candidate for
the detection of $\cal MSSM$ Higgs bosons at large values of tan$\beta$.\par
We have not made a complete study of reactions
(\ref{e1})--(\ref{e3}) at LHC, but we have checked in a few cases that the
usual
$6\div 10$ reduction factor applies to the cross sections.

\subsection*{Conclusions}
In this paper we have studied,
in the $\cal MSSM$, a number of processes for
the production of a neutral, intermediate mass Higgs boson
with additional heavy particles in the final state which
can be used for tagging purposes.
We find that, for large values of tan$\beta$, the cross sections
for $pp\rightarrow bZ^0\Phi$ are of the order of 10 $pb$ or more,
over the whole intemediate range of $M_{A^0}$, for $A^0$ and at least one of
the other two Higgses.

\newpage

\vspace*{\fill}

\subsection*{Table Captions}
\begin{description}

\item[table I  ] ${\cal MSSM}$ neutral Higgs couplings
to the massive fermions $\tau,b$ and $t$, and to the massive gauge
bosons $W^\pm$ and $Z^0$.

\end{description}

\vspace*{\fill}

\subsection*{Figure Captions}
\begin{description}
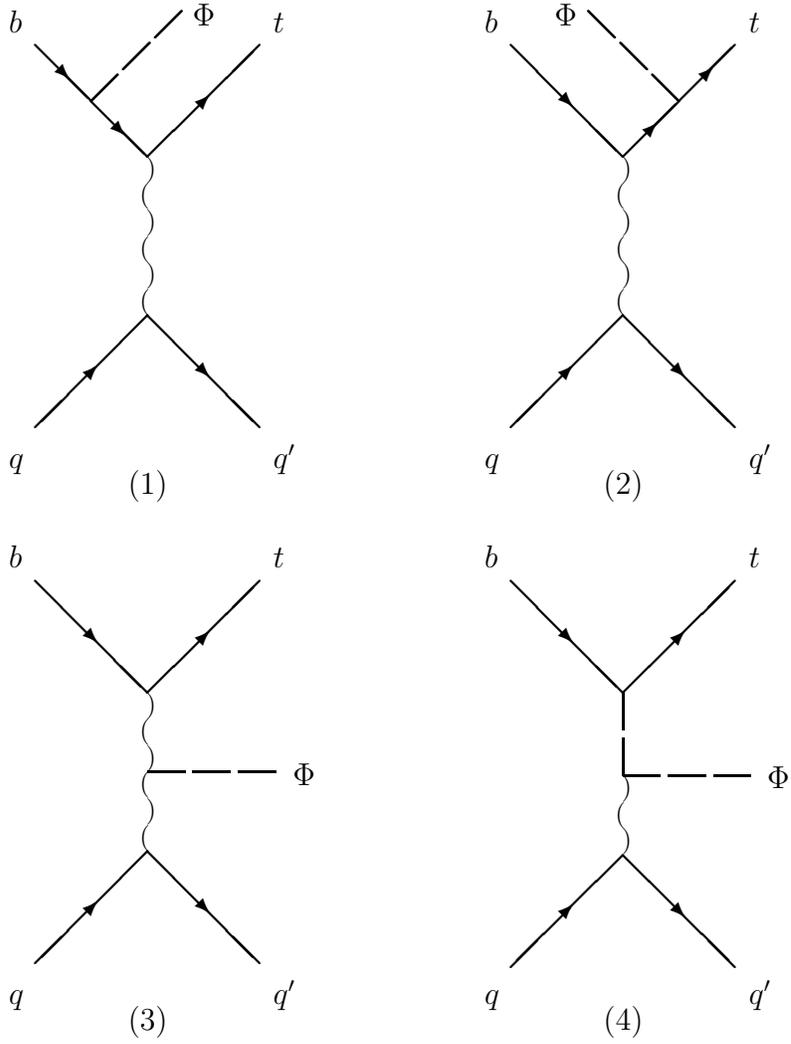
\item[fig.1 ] Feynman diagrams contributing in the lowest order to
$bq\rightarrow t q' \Phi$, where $q,q'=u,d,s,c$ and
$\Phi=\phi,H^0,h^0,A^0$, as appropriate.
For the ${\cal SM}$ Higgs ($\Phi=\phi$) only the first three
contribute, while in the ${\cal MSSM}$ case,
for a $CP$--odd Higgs boson ($\Phi=A^0$), diagram (3) is absent.

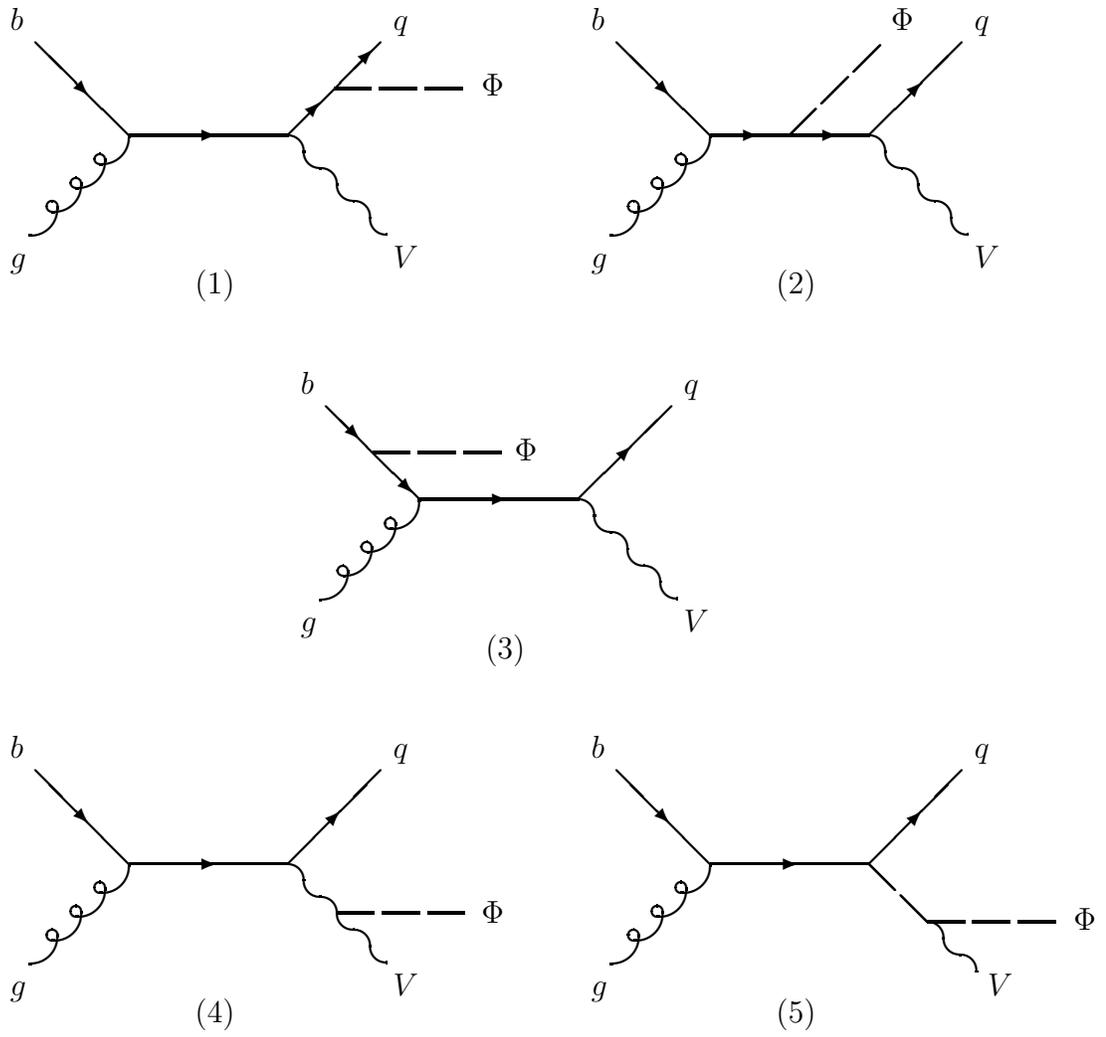
\item[fig.2 ] Basic Feynman diagrams contributing in the lowest order to
$bg\rightarrow q V \Phi$, where $q=b,t$; $V=W^-,Z^0$ and $\Phi=\phi,H^0,h^0,
A^0$, as appropriate. For the ${\cal SM}$ Higgs ($\Phi=\phi$)
only the first four contribute, while in the ${\cal MSSM}$ case,
for a $CP$--odd Higgs boson ($\Phi=A^0$), diagram (4) is absent.

\item[fig.3 ] Cross sections for a number of processes
$b+q(\bar q)\rightarrow t+q'(\bar q')+\Phi$,
$b+g\rightarrow t+W^-+\Phi$ and
$b+g\rightarrow b+Z^0+\Phi$ plus
their charge conjugated at SSC for $m_t=150$ GeV
and tan$\beta = 2$.
Each curve is labeled with the name of the
Higgs boson it refers to.

\item[fig.4 ] Cross sections of the processes
$b+q(\bar q)\rightarrow t+q'(\bar q')+\Phi^0$,
$\Phi = h, H, A$ and
their charge conjugated at SSC for tan$\beta = 15$
(lower curves) and tan$\beta = 30$ (upper curves).
The $top$ mass is 150 GeV.

\item[fig.5 ] Cross sections of the processes
$b+g\rightarrow t+W^-+\Phi^0$,
$\Phi = h, H, A$ and
their charge conjugated at SSC for tan$\beta = 15$
(lower curves) and tan$\beta = 30$ (upper curves).
The $top$ mass is 150 GeV.

\item[fig.6 ] Cross sections of the processes
$b+g\rightarrow b+Z^0+\Phi^0$,
$\Phi = h, H, A$ and
their charge conjugated at SSC for tan$\beta = 15$
(lower curves) and tan$\beta = 30$ (upper curves).

\end{description}

\vspace*{\fill}

\newpage
\setcounter{page}{0}

\begin{table}%[p]%[htbp]
\begin{center}
\begin{tabular}{|c|c|c|c|}     \hline
\rule[-0.6cm]{0cm}{1.3cm}
$\;\;\;\;\;$ & $h^0$ & $H^0$ & $A^0$  \\ \hline
\rule[-0.6cm]{0cm}{1.3cm}
$t \bar{t}$ & $\frac{\cos \alpha}{\sin \beta}$ & $\frac{\sin \alpha}
{\sin \beta}$ & $-i\gamma _{5} \cot \beta$   \\ %\hline
\rule[-0.6cm]{0cm}{1.3cm}
$b \bar{b}$ , $\tau \bar{\tau}$ & $-\frac{\sin \alpha}{\cos \beta}$
& $\frac{\cos \alpha}{\cos \beta}$ & $-i\gamma _{5} \tan \beta$ \\ %\hline
\rule[-0.6cm]{0cm}{1.3cm}
$W^\pm W^\mp,Z^0Z^0$ & $\sin (\beta - \alpha)$ & $\cos(\beta - \alpha)$ & $0$
\\ %\hline
\rule[-0.6cm]{0cm}{1.3cm}
$H^{\pm} W^\mp$ & $\cos(\alpha - \beta)$ & $\sin(\alpha - \beta)$ & $1$    \\
%\hline
\rule[-0.6cm]{0cm}{1.3cm}
$A^0Z^0$ & $\cos(\alpha - \beta)$ & $\sin(\alpha - \beta)$ & $0$    \\ \hline
\end{tabular}
\end{center}
\end{table}

\centerline{\Large Table I}
\vfill

\newpage
\setcounter{page}{0}

\
\vskip2.0cm

\begin{picture}(34500,14500)
\THICKLINES
\drawline\fermion[\SE\REG](7000,15000)[3000]
\drawarrow[\LDIR\ATBASE](\pmidx,\pmidy)
\drawline\fermion[\SE\REG](\fermionbackx,\fermionbacky)[3000]
\drawarrow[\LDIR\ATBASE](\pmidx,\pmidy)
\seglength=1416  \gaplength=300  % Changes the \scalar defaults.
\drawline\scalar[\NE\REG](\fermionfrontx,\fermionfronty)[3]
\drawline\fermion[\NE\REG](\fermionbackx,\fermionbacky)[6000]
\drawarrow[\LDIR\ATBASE](\pmidx,\pmidy)
\bigphotons
\drawline\photon[\S\REG](\fermionfrontx,\fermionfronty)[6]
\drawline\fermion[\SW\REG](\pbackx,\pbacky)[6000]
\drawarrow[\NE\ATBASE](\pmidx,\pmidy)
\drawline\fermion[\SE\REG](\pfrontx,\pfronty)[6000]
\drawarrow[\LDIR\ATBASE](\pmidx,\pmidy)
\put(6000,15500){$b$}
\put(16000,15500){$t$}
\put(12950,15750){$\Phi$}
\put(6000,-1000){$q$}
\put(16000,-1000){$q'$}
\put(10500,-2000){$(1)$}
\drawline\fermion[\SE\REG](25000,15000)[6000]
\drawarrow[\LDIR\ATBASE](\pmidx,\pmidy)
\bigphotons
\drawline\photon[\S\REG](\fermionbackx,\fermionbacky)[6]
\drawline\fermion[\NE\REG](\fermionbackx,\fermionbacky)[3000]
\drawarrow[\LDIR\ATBASE](\pmidx,\pmidy)
\drawline\fermion[\NE\REG](\fermionbackx,\fermionbacky)[3000]
\drawarrow[\LDIR\ATBASE](\pmidx,\pmidy)
\seglength=1416  \gaplength=300  % Changes the \scalar defaults.
\drawline\scalar[\NW\REG](\fermionfrontx,\fermionfronty)[3]
\drawline\fermion[\SW\REG](\photonbackx,\photonbacky)[6000]
\drawarrow[\NE\ATBASE](\pmidx,\pmidy)
\drawline\fermion[\SE\REG](\pfrontx,\pfronty)[6000]
\drawarrow[\LDIR\ATBASE](\pmidx,\pmidy)
\put(24000,15500){$b$}
\put(34000,15500){$t$}
\put(26650,15750){$\Phi$}
\put(24000,-1000){$q$}
\put(34000,-1000){$q'$}
\put(28500,-2000){$(2)$}
\end{picture}

\vspace{2.0cm}

\begin{picture}(34500,14500)
\THICKLINES
\drawline\fermion[\SE\REG](7000,15000)[6000]
\drawarrow[\LDIR\ATBASE](\pmidx,\pmidy)
\drawline\fermion[\NE\REG](\fermionbackx,\fermionbacky)[\fermionlength]
\drawarrow[\LDIR\ATBASE](\pmidx,\pmidy)
\bigphotons
\drawline\photon[\S\REG](\fermionfrontx,\fermionfronty)[6]
\seglength=1416  \gaplength=300  % Changes the \scalar defaults.
\drawline\scalar[\E\REG](\pmidx,\pmidy)[3]
\drawline\fermion[\SW\REG](\photonbackx,\photonbacky)[\fermionlength]
\drawarrow[\NE\ATBASE](\pmidx,\pmidy)
\drawline\fermion[\SE\REG](\pfrontx,\pfronty)[\fermionlength]
\drawarrow[\LDIR\ATBASE](\pmidx,\pmidy)
\put(6000,15500){$b$}
\put(16000,15500){$t$}
\put(16750,7250){$\Phi$}
\put(6000,-1000){$q$}
\put(16000,-1000){$q'$}
\put(10500,-2000){$(3)$}
\drawline\fermion[\SE\REG](25000,15000)[6000]
\drawarrow[\LDIR\ATBASE](\pmidx,\pmidy)
\drawline\fermion[\NE\REG](\fermionbackx,\fermionbacky)[\fermionlength]
\drawarrow[\LDIR\ATBASE](\pmidx,\pmidy)
\seglength=1416  \gaplength=300  % Changes the \scalar defaults.
\drawline\scalar[\S\REG](\fermionfrontx,\fermionfronty)[2]
\seglength=1416  \gaplength=300  % Changes the \scalar defaults.
\drawline\scalar[\E\REG](\scalarbackx,\scalarbacky)[3]
\bigphotons
\drawline\photon[\S\REG](\pfrontx,\pfronty)[3]
\drawline\fermion[\SW\REG](\photonbackx,\photonbacky)[\fermionlength]
\drawarrow[\NE\ATBASE](\pmidx,\pmidy)
\drawline\fermion[\SE\REG](\pfrontx,\pfronty)[\fermionlength]
\drawarrow[\LDIR\ATBASE](\pmidx,\pmidy)
\put(24000,15500){$b$}
\put(34000,15500){$t$}
\put(34700,7150){$\Phi$}
\put(24000,-1000){$q$}
\put(34000,-1000){$q'$}
\put(28500,-2000){$(4)$}
\end{picture}

\vspace{2.0cm}

\centerline{\bf\Large Fig.1}

\vfill
\newpage
\setcounter{page}{0}

\
\vskip2.0cm

\begin{picture}(10000,8000)
\THICKLINES
\bigphotons
\drawline\fermion[\W\REG](12000,8000)[6000]
\drawarrow[\E\ATBASE](\pmidx,\pmidy)
\drawline\fermion[\NW\REG](\fermionbackx,\fermionbacky)[5000]
\drawarrow[\SE\ATBASE](\pmidx,\pmidy)
\drawline\gluon[\SW\REG](6000,8000)[3]
\drawline\fermion[\NE\REG](12000,8000)[5000]
\drawarrow[\NE\ATBASE](13000,9000)
\drawarrow[\NE\ATBASE](15000,11000)
\seglength=1416  \gaplength=300  % Changes the \scalar defaults.
\drawline\scalar[\E\REG](\pmidx,\pmidy)[3]
\drawline\photon[\SE\REG](12000,8000)[6]
\put(1500,12000){$b$}
\put(1500,3000){$g$}
\put(16000,12000){$q$}
\put(16000,3000){$V$}
\put(19350,9500){$\Phi$}
\put(8500,2000){$(1)$}
\drawline\fermion[\W\REG](34000,8000)[6000]
\drawarrow[\E\ATBASE](32500,8000)
\drawarrow[\E\ATBASE](29500,8000)
\seglength=1416  \gaplength=300  % Changes the \scalar defaults.
\drawline\scalar[\NE\REG](\pmidx,\pmidy)[3]
\drawline\fermion[\NW\REG](\fermionbackx,\fermionbacky)[5000]
\drawarrow[\SE\ATBASE](\pmidx,\pmidy)
\drawline\gluon[\SW\REG](28000,8000)[3]
\drawline\fermion[\NE\REG](34000,8000)[5000]
\drawarrow[\NE\ATBASE](\pmidx,\pmidy)
\drawline\photon[\SE\REG](34000,8000)[6]
\put(23500,12000){$b$}
\put(23500,3000){$g$}
\put(38000,12000){$q$}
\put(38000,3000){$V$}
\put(34850,12000){$\Phi$}
\put(30500,2000){$(2)$}
\end{picture}

\vskip 2.0cm

\begin{picture}(21000,8000)
\THICKLINES
\bigphotons
\drawline\fermion[\W\REG](23000,8000)[6000]
\drawarrow[\E\ATBASE](\pmidx,\pmidy)
\drawline\fermion[\NW\REG](\fermionbackx,\fermionbacky)[5000]
\drawarrow[\SE\ATBASE](16500,8500)
\drawarrow[\SE\ATBASE](14500,10500)
\seglength=1416  \gaplength=300  % Changes the \scalar defaults.
\drawline\scalar[\E\REG](\pmidx,\pmidy)[3]
\drawline\gluon[\SW\REG](17000,8000)[3]
\drawline\fermion[\NE\REG](23000,8000)[5000]
\drawarrow[\NE\ATBASE](\pmidx,\pmidy)
\drawline\photon[\SE\REG](23000,8000)[6]
\put(12500,12000){$b$}
\put(12500,3000){$g$}
\put(27000,12000){$q$}
\put(27000,3000){$V$}
\put(20600,9500){$\Phi$}
\put(19500,2000){$(3)$}
\end{picture}

\vskip 2.0cm

\begin{picture}(10000,8000)
\THICKLINES
\bigphotons
\drawline\fermion[\W\REG](12000,8000)[6000]
\drawarrow[\E\ATBASE](\pmidx,\pmidy)
\drawline\fermion[\NW\REG](\fermionbackx,\fermionbacky)[5000]
\drawarrow[\SE\ATBASE](\pmidx,\pmidy)
\drawline\gluon[\SW\REG](6000,8000)[3]
\drawline\fermion[\NE\REG](12000,8000)[5000]
\drawarrow[\NE\ATBASE](\pmidx,\pmidy)
\drawline\photon[\SE\REG](12000,8000)[6]
\seglength=1416  \gaplength=300  % Changes the \scalar defaults.
\drawline\scalar[\E\REG](\pmidx,\pmidy)[3]
\put(1500,12000){$b$}
\put(1500,3000){$g$}
\put(16000,12000){$q$}
\put(16000,3000){$V$}
\put(19350,5750){$\Phi$}
\put(8500,2000){$(4)$}
\drawline\fermion[\W\REG](34000,8000)[6000]
\drawarrow[\E\ATBASE](\pmidx,\pmidy)
\drawline\fermion[\NW\REG](\fermionbackx,\fermionbacky)[5000]
\drawarrow[\SE\ATBASE](\pmidx,\pmidy)
\drawline\gluon[\SW\REG](28000,8000)[3]
\drawline\fermion[\NE\REG](34000,8000)[5000]
\drawarrow[\NE\ATBASE](\pmidx,\pmidy)
\seglength=1416  \gaplength=300  % Changes the \scalar defaults.
\drawline\scalar[\SE\REG](34000,8000)[2]
\drawline\photon[\SE\REG](\scalarbackx,\scalarbacky)[3]
\seglength=1416  \gaplength=300  % Changes the \scalar defaults.
\drawline\scalar[\E\REG](\scalarbackx,\scalarbacky)[3]
\put(23500,12000){$b$}
\put(23500,3000){$g$}
\put(38000,12000){$q$}
\put(38500,3000){$V$}
\put(41750,5500){$\Phi$}
\put(30500,2000){$(5)$}
\end{picture}

\vskip 2.0cm

\centerline{\bf\Large Fig.2}
\vfill

\end{document}